\documentclass[runningheads]{llncs}
\usepackage[T1]{fontenc}
\usepackage[table]{xcolor}
\usepackage{colortbl}
\usepackage{graphicx}
\usepackage{amsmath}
\usepackage{amssymb}
\usepackage{algorithm}
\usepackage{algpseudocode}
\usepackage{pgfplots}
\usepackage{tikz}
\usepackage{amsmath}
\usepackage{orcidlink}
\pgfplotsset{compat=1.18} 
\usetikzlibrary{shapes} 
\usetikzlibrary{calc}
\usepackage{enumitem}
\usepackage{multirow}
\usepackage{textcomp}
\usepackage{subcaption}
\usepackage{tabu}
\usepackage{color}
\usepackage{url}
\usepackage{hyperref}
\usepackage{breakurl}

\begin{document}
\title{Lost in Models? Structuring Managerial Decision Support in Process Mining with Multi-criteria Decision Making}
\titlerunning{Managerial Decision Support in Process Mining with Multi-criteria Decision}
\author{Rob H. Bemthuis\,\orcidlink{0000-0003-2791-6070}
}
\authorrunning{R. H. Bemthuis}
\institute{University of Twente, Enschede, The Netherlands\\
\email{r.h.bemthuis@utwente.nl}}
\maketitle              
\begin{abstract}
Process mining is increasingly adopted in modern organizations, producing numerous process models that, while valuable, can lead to model overload and decision-making complexity. This paper explores a multi-criteria decision-making (MCDM) approach to evaluate and prioritize process models by incorporating both quantitative metrics (e.g., fitness, precision) and qualitative factors (e.g., cultural fit). An illustrative logistics example demonstrates how MCDM, specifically the Analytic Hierarchy Process (AHP), facilitates trade-off analysis and promotes alignment with managerial objectives. Initial insights suggest that the MCDM approach enhances context-sensitive decision-making, as selected models address both operational metrics and broader managerial needs. While this study is an early-stage exploration, it provides an initial foundation for deeper exploration of MCDM-driven strategies to enhance the role of process mining in complex organizational settings. 

\keywords{Process Mining \and Multi-criteria Decision Making \and Managerial Decision Support \and Analytical Hierarchy Process}
\end{abstract}

\section{Introduction}
\label{section:introduction}
Recent advances in process mining have improved the ability to capture and analyze complex organizational workflows through event logs. However, this progress has led to an increasing abundance of process models, often overlapping in scope or providing divergent insights for different stakeholders (e.g., operational vs.\ managerial)~\cite{batista2018process,maneschijn2022methodology,maneschijn2022balancing}. This “model overload” phenomenon presents strategic challenges: rather than supporting decision-making, the sheer volume of (apparently) competing models can obscure key insights and hamper the alignment of process analytics with organizational objectives. As a result, managers may struggle to distinguish relevant models from irrelevant ones, making it difficult to focus on actionable insights (see e.g.,~\cite{akhramovich2024systematic,augusto2022connection,elkhovskaya2023extending,imran2022complex}). 

While process mining accelerates business process digitalization, its effectiveness depends on deeper integration with organizational goals, key performance indicators, and managerial expertise~\cite{reinkemeyer2020process,vom2021five}. Achieving this integration across diverse processes and stakeholders necessitates a robust decision support mechanism that balances tacit knowledge with empirical findings. Traditional decision support systems often fall short in this regard, as they struggle to incorporate subjective managerial perspectives alongside quantitative process metrics. Consequently, decision-makers must navigate complex model repositories without structured guidance, increasing the risk of suboptimal or misaligned choices. 

Multi-Criteria Decision-Making (MCDM) provides a structured framework for evaluating and prioritizing process models by combining quantitative performance metrics with qualitative managerial insights~\cite{koksalan2011multiple}. As a well-established decision analysis method, MCDM is particularly effective when no single optimal solution exists, enabling decision-makers to navigate trade-offs between competing criteria~\cite{aruldoss2013survey,sahoo2023comprehensive}. Applying MCDM to process mining extends beyond evaluating models based solely on fitness or precision, promoting alignment with both strategic and operational objectives. 

This paper presents an approach for applying MCDM to address model overload in process mining. We propose an approach that synthesizes process mining outputs while integrating an organization’s strategic priorities. By combining objective indicators (e.g., fitness, precision) with managerial assessments, MCDM provides a structured way for evaluating and prioritizing process models. This approach offers two key benefits: aligning model selection with strategic objectives and clarifying trade-offs in multi-stakeholder decision-making.

To illustrate the potential of this approach, we present an illustrative example in a logistics context, where the Analytic Hierarchy Process (AHP) is applied as an MCDM approach to filter and prioritize mined process models. Initial findings suggest that MCDM-based methods can reduce model selection complexity, guide resource allocation toward high-impact analyses, and improve communication between technical and managerial stakeholders. While full-scale validation remains an area for future research, we expect that this concept will generate valuable discussions on integrating decision-making theory, managerial insights, and process mining methods. This work proposes a structured approach for more strategic and context-aware use of process models. 

The remainder of this paper is as follows. Section~\ref{section:related work} discusses related work. Section~\ref{section:approach} introduces the proposed MCDM approach. In Sect.~\ref{section:illustrative example}, we present an illustrative example in the logistics domain, highlighting initial findings and challenges. Section~\ref{section:conclusions} concludes the paper.

\section{Related Work}
\label{section:related work}
The growing number of discovered process models often results in highly complex structures, commonly referred to as “spaghetti models”, and an overwhelming number of variations. These models can obscure meaningful patterns, making it difficult to derive actionable insights~\cite{klessascheck2021domain}. While filtering and abstraction techniques help manage complexity, they frequently introduce redundant or conflicting perspectives, further complicating their alignment with organizational objectives~\cite{maneschijn2022balancing}. Moreover, beyond the complexity of each individual model, the sheer volume of potentially overlapping or incompatible models can compound decision-making challenges. Research highlights that stakeholder needs contribute to model overload: highly detailed models, though technically accurate, may be too complex for managerial decision-making~\cite{ammann2025process}. Analysts often struggle to determine which variant best represents the process in a given context, further complicating their evaluations. 

Several techniques have been proposed to address complexity. Existing solutions focus on model filtering~\cite{conforti2016filtering}, abstraction~\cite{jagadeesh2009abstractions,van2021event}, and domain-specific metrics~\cite{hidalgo2022domain}. Trace clustering~\cite{song2009trace} is a well-known method that groups similar traces and removes minor variations to enhance process comprehension. However, finding the right level of abstraction remains a challenge: excessive simplification may obscure critical details, while insufficient abstraction leaves models too complex. Process performance metrics, such as fitness, precision, and generalization~\cite{buijs2012role}, can assist in selecting promising process models. Furthermore, complexity measures—including control-flow complexity and node/edge counts—help detect overly complex models~\cite{augusto2022connection}. Nonetheless, these techniques primarily enhance structural clarity rather than directly supporting strategic decision-making. 

Despite advancements in process mining, consolidating multiple discovered models into a coherent, decision-driven framework remains challenging. Most approaches emphasize structural refinement but often neglect managerial preferences, which favor simplicity and strategic relevance over purely technical model quality. Recent studies have explored MCDM techniques in process mining. For example, \cite{dos2025smart} applied MCDM to rank industrial machines for maintenance planning, integrating technical indicators with expert judgment. Similarly, \cite{dogan2021process} used AHP for process mining technology selection, incorporating uncertainty and sensitivity analysis to improve ranking robustness. While these studies demonstrate MCDM’s potential for process-related decision-making, they focus on technology and asset selection rather than process model evaluation. Our work builds on that foundation by applying MCDM, specifically AHP, to structurally compare and prioritize discovered process models, thereby contributing to alignment with managerial objectives.

\section{Proposed Approach}
\label{section:approach}
This section presents an MCDM approach to assist in selecting and prioritizing process models (typically obtained from large repositories). As illustrated in Fig.~\ref{fig:approach}, the approach structures decision-making by ranking and selecting models based on multiple, potentially conflicting criteria: 

\begin{figure}[ht]
    \centering
    \includegraphics[width=0.76\linewidth]{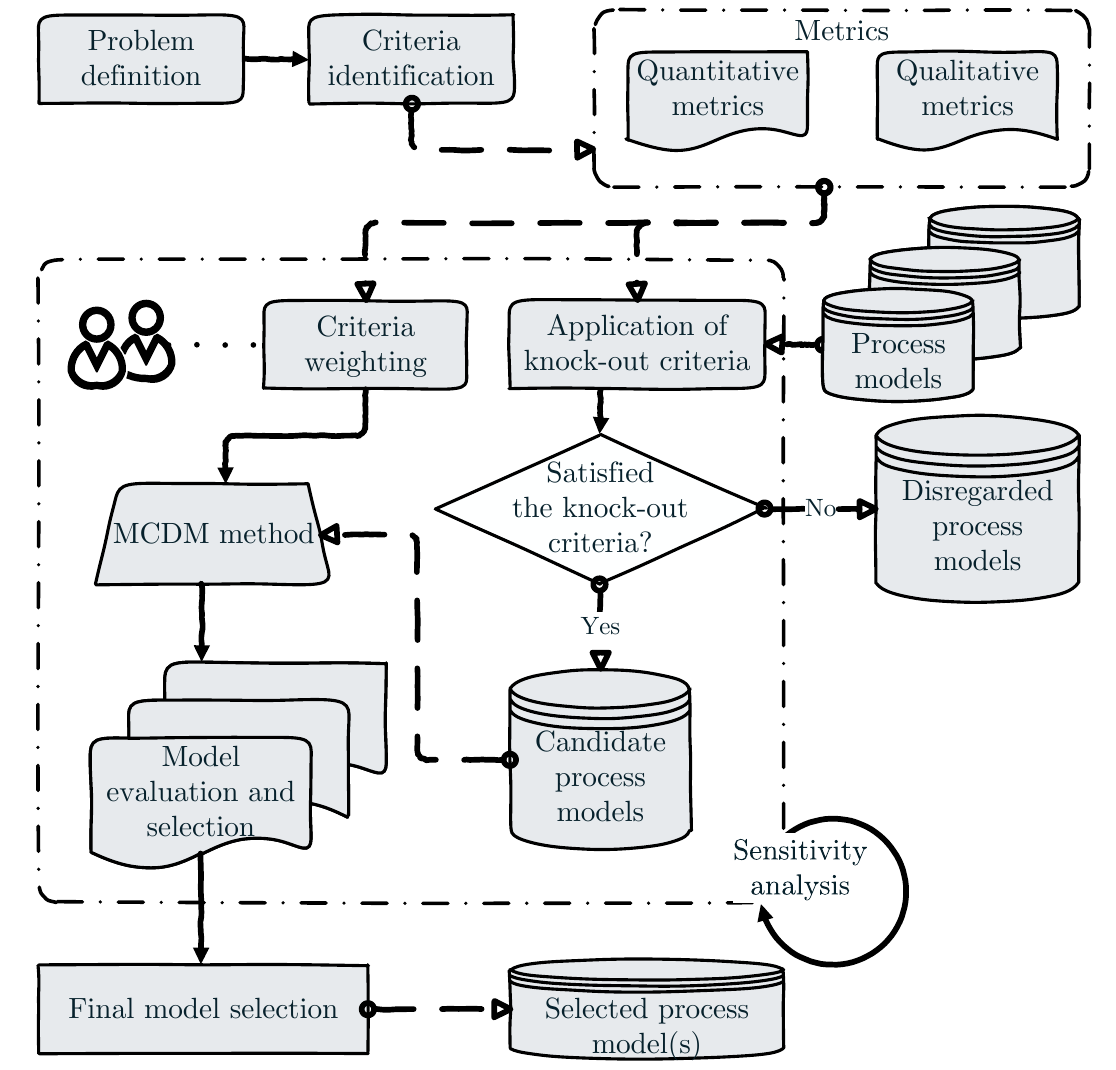}
    \caption{Proposed MCDM approach for process model selection.}
    \label{fig:approach}
\end{figure}

\begin{enumerate}
    \item \textbf{Problem definition}: define the selection objective, identifying the most relevant process model(s) from a set of discovered alternatives. 
    
    \item \textbf{Criteria identification}: evaluation relies on two broad categories.         \textbf{Quantitative metrics}: process mining measures, such as fitness, precision, generalization, and simplicity. \textbf{Qualitative metrics}: managerial factors, such as decision-support value (e.g., the model’s ability to highlight operational inefficiencies), stakeholder alignment (e.g., relevance to different stakeholder groups), and implementation feasibility (e.g., potential impact). 
    \item \textbf{Application of knock-out criteria}: models failing key constraints—such as structural completeness, data quality, or regulatory compliance—are eliminated early. 
    \item \textbf{Criteria weighting}: weights are determined using pairwise comparisons, entropy-based weighting, or expert input, to reflect each criterion’s relative importance. 
    \item \textbf{Model evaluation and selection}: process models are ranked using an MCDM technique. The choice of method depends on the decision context, data availability, and stakeholder preferences (see \cite{zavadskas2014state} for an overview). 
    \item \textbf{Sensitivity analysis}: criteria weights are varied to assess ranking stability under different scenarios. 
\end{enumerate}

By structuring the selection process in these steps, this approach can mitigate bias, reduce reliance on purely technical measures, and better align process mining outputs with managerial decision-making. 

\section{Illustrative Example}
\label{section:illustrative example}
To illustrate the feasibility of our proposed MCDM approach, we apply it to a logistics case study presented in~\cite{bemthuis2019agent}, which focuses on selecting process models derived from event logs. The study includes a dataset comprising 270 event logs generated across 27 distinct system configurations~\cite{edocwData}, with each configuration yielding 20 log files. For this illustrative example, we concentrate on the first event log per experiment, thus evaluating one event log per configuration. Process models were extracted using the Inductive Miner, which is expected to produce sound, relatively simple models, and were subsequently evaluated.

\subsection{Problem Definition}
The objective of this illustrative study is to support decision-makers in selecting the most suitable \emph{configuration for investment}, taking into account uncertainty in the decision-making process. Process models function as part of multiple evaluation criteria, alongside throughput time and implementation risk. A key challenge is to balance technical accuracy with practical feasibility, thereby contributing to a more informed and strategic investment decision. 

\subsection{Criteria Identification}
\begin{figure*}[ht]
\centering
\begin{tikzpicture}
	\begin{axis}[%
	scatter/classes={%
		a={mark=x,blue},
		b={mark=triangle,red},
		c={mark=o,green!60!black},
		d={mark=oplus,mark options={scale=0.8},mark size=3pt,blue!50!red}},
		y=4.5cm,
		x={0.0355*\textwidth},
		legend style={at={(0.5,-0.6)},anchor=north,legend cell align=left},
		legend columns=4,
	    symbolic x coords={411,412,413,421,422,423,431,432,433,511,512,513,521,522,523,531,532,533,611,612,613,621,622,623,631,632,633},
	    xtick={411,412,413,421,422,423,431,432,433,511,512,513,521,522,523,531,532,533,611,612,613,621,622,623,631,632,633},
	    ylabel={Criteria score},
	    y label style={at={(axis description cs:-0.07,0.5)},anchor=north},
	    xlabel={Configuration as specified in \cite{bemthuis2019agent}},
	    x label style={at={(axis description cs:0.5,-0.18)},anchor=north}, 
	    ymin=0.75,
	    ymax=1.05,
	    enlarge x limits={abs=5pt},
	    font=\tiny, 
	]
	\addplot[scatter,only marks,%
		scatter src=explicit symbolic]%
	table[meta=label] {
x     y      label
411	0.999546682	a
412	0.999950642	a
413	0.999741972	a
421	0.994365331	a
422	0.999754702	a
423	0.993002813	a
431	0.984544199	a
432	0.984384268	a
433	0.998490908	a
511	0.989859649	a
512	0.996299226	a
513	0.983236842	a
521	0.987025983	a
522	0.990047435	a
523	0.973798246	a
531	0.996209224	a
532	0.999671544	a
533	0.988753388	a
611	0.987359748	a
612	0.98970021	a
613	0.971592502	a
621	0.988437938	a
622	0.991070175	a
623	0.968894737	a
631	0.995129812	a
632	0.996677772	a
633	0.977897599	a
411	0.79968	b
412	0.79861	b
413	0.79906	b
421	0.79896	b
422	0.8	b
423	0.79957	b
431	0.82558	b
432	0.82493	b
433	0.83534	b
511	0.79918	b
512	0.79955	b
513	0.79874	b
521	0.79899	b
522	0.79906	b
523	0.79893	b
531	0.83855	b
532	0.8379	b
533	0.83844	b
611	0.79899	b
612	0.79924	b
613	0.79893	b
621	0.79906	b
622	0.79887	b
623	0.79882	b
631	0.86162	b
632	0.86299	b
633	0.83496	b
411	0.98135	c
412	0.89924	c
413	0.97027	c
421	0.99598	c
422	0.96416	c
423	0.99659	c
431	1	c
432	0.99999	c
433	1	c
511	0.99729	c
512	0.99523	c
513	0.99799	c
521	0.99795	c
522	0.99705	c
523	0.99849	c
531	1	c
532	1	c
533	1	c
611	0.99795	c
612	0.99668	c
613	0.99849	c
621	0.99705	c
622	0.99779	c
623	0.9986	c
631	1	c
632	1	c
633	1	c
	};

\legend{\scriptsize{Fitness}, \scriptsize{Precision},\scriptsize{Generalization}}
	\end{axis}
\end{tikzpicture}
\caption{Evaluation of process models extracted with the Inductive Miner.}
\label{fig:inductive miner}
\end{figure*}
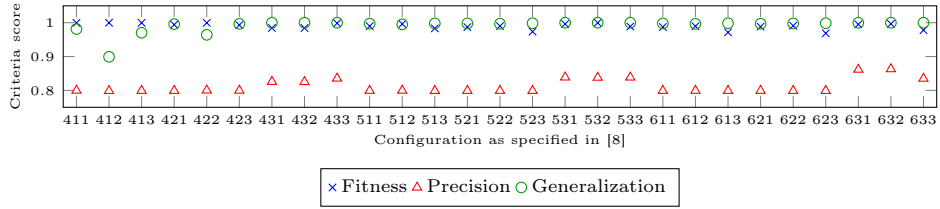

To evaluate model quality, we consider quantitative metrics associated with key process mining quality dimensions: fitness, precision, and generalization. Simplicity is excluded due to its strong correlation with generalization~\cite{buijs2014quality}. Using the Inductive Miner, we obtained the scores for the scenarios detailed in~\cite{bemthuis2019agent}, as illustrated in Fig.~\ref{fig:inductive miner}. As additional criteria, we include the throughput times specified in~\cite{bemthuis2019agent} and the implementation risk linked to business goal alignment. 

\subsection{Application of Knock-Out Criteria}
To ensure high-quality model selection, we set a strict fitness threshold of $0.999$. Out of the initial $27$ process models, only $5$ meet this criterion and are retained for further analysis. 

\subsection{Criteria Weighting}
We use Saaty's AHP method~\cite{saaty2008decision}, which is widely applied in decision-making~\cite{aziz2016mcdm}, to determine the relative importance of \textit{fitness} ($F$), \textit{precision} ($P$), and \textit{generalization} ($G$) via expert pairwise comparisons (Table~\ref{table:pairwise comparisons}). The resulting weights are $w_F = 0.57$, $w_P = 0.22$, and $w_G = 0.21$. \textit{Throughput time} ($T$) is categorized into low ($0$–$50$ min, $C_2 = 1.0$), medium ($50$–$100$ min, $C_2 = 0.75$), and high ($>100$ min, $C_2 = 0.50$). \textit{Implementation risk} ($IR$), assessed externally, is classified as low ($C_3 = 1.0$), medium ($C_3 = 0.70$), or high ($C_3 = 0.50$). The \textit{overall weight allocation} is $w_1 = 0.40$ (process model quality), $w_2 = 0.25$ (throughput time), and $w_3 = 0.35$ (implementation risk).

\begin{table}[ht]
\caption{Pairwise comparison results for $F$, $P$, and $G$.}
\label{table:pairwise comparisons}
\centering
    \begin{tabular}{|l|llll|llll|llll|}
    \hline
    \multicolumn{1}{|l|}{Criteria} & \multicolumn{4}{c|}{\textbf{Stakeholder 1}} & \multicolumn{4}{c|}{\textbf{Stakeholder 2}} & \multicolumn{4}{c|}{\textbf{Stakeholder 3}} \\
    \hline
     & $F$ & $P$ & $G$ & Weight 
     & $F$ & $P$ & $G$ & Weight 
     & $F$ & $P$ & $G$ & Weight \\
    \hline
    $F$ & -  & $6.00$  & $7.00$ & $0.76$ & - & $5.00$ & $5.00$ & $0.71$ & - & $1.0$ & $0.33$ & $0.23$\\
    $P$ & $0.17$  & -   & $1.00$  & $0.12$ & $0.20$ & - & $1.00$ & $0.14$ & $1.00$ & - & $2$ & $0.40$\\
    $G$ & $0.14$  & $1.00$ & - & $0.12$ & $0.20$ & $1.00$ & - & $0.14$ & $3.00$ & $0.50$ & - & $0.37$\\
    \hline
    \end{tabular}
\end{table}

\subsection{Model Evaluation and Selection}
\label{subsection:model evaluation}
The \textit{final model score} is computed as: $C_{\text{total}} = \sum_{i=1}^{n} w_i C_i$, where $C_1$ (process model quality), $C_2$ (throughput time), and $C_3$ (implementation risk) contribute to the ranking (see Table~\ref{tab:config_comparison}). Although 532 leads in $C_1$ and $C_2$, 411 ranks best overall due to its balanced performance. This highlights the importance of multi-criteria evaluation integrating technical quality and practical feasibility. 

\begin{table}[ht]
    \caption{Performance comparison of different configurations.}
    \label{tab:config_comparison}
    \centering
    \begin{tabular}{|c|c|c|c|>{\columncolor[gray]{0.9}}c|c|>{\columncolor[gray]{0.9}}c|>{\columncolor[gray]{0.9}}c|>{\columncolor[gray]{0.7}}c|}
        \hline
        Conf.
        & $F$ & $P$ & $G$
        & $C_1$
        & $T$ (min)
        & $C_2$ 
        & $C_3$
        & $C_{\text{total}}$
        \\
        \hline
        411 & 1.000 & 0.800 & 0.981 & 0.952 & 73.3 & 0.70 (medium) & 0.75 (medium) & \textbf{0.818} \\
        412 & 1.000 & 0.799 & 0.899 & 0.934 & 211.9 & 0.50 (high) & 0.75 (medium) & 0.761 \\
        413 & 1.000 & 0.799 & 0.970 & 0.949 & 74.3 & 0.70 (medium) & 0.75 (medium) & 0.817 \\
        422 & 1.000 & 0.800 & 0.964 & 0.948 & 107.3 & 0.50 (high) & 0.75 (medium) & 0.767 \\
        532 & 1.000 & 0.825 & 1.000 & 0.964 & 33.13 & 1.0 (low) & 0.50 (high) & 0.811 \\
        \hline
    \end{tabular}
\end{table}

Although we do not include a sensitivity analysis in this study, future work could explore its effect by adjusting criteria weights to assess ranking robustness. Further validation may involve expanding beyond traditional process mining dimensions (e.g., fitness, precision, generalization) to incorporate additional metrics, both quantitative and qualitative, as well as experimenting with alternative MCDM methods to enhance model selection stability. 

\section{Conclusions}
\label{section:conclusions}
This study highlights the potential of MCDM as a structured approach for mitigating model overload in process mining. By integrating quantitative metrics with managerial considerations, the proposed approach enables more context-sensitive selection and prioritization of process models. The logistics-based illustrative example demonstrates how MCDM facilitates trade-off analysis and promotes alignment between technical criteria and managerial objectives. 

While the approach shows promise, further research is needed to evaluate alternative MCDM methods, incorporate additional decision criteria (e.g., risk, cost, compliance), and expand sensitivity analysis to assess ranking stability. Real-world adoption will also require addressing potential resistance to using the approach among stakeholders. Integration with existing process mining tools may further enhance practical applicability. Ultimately, empirical validation remains key to refining this approach into a broadly applicable decision-support solution.

\begin{credits}
\subsubsection{\ackname} This work was (financially) supported by the Dutch Ministry of Infrastructure and Water Management and TKI Dinalog (ECOLOGIC project; case no. 31192090). 

\end{credits}

\bibliographystyle{splncs04}
\bibliography{mybibliography}
\end{document}